\documentclass[proof]{WileyASNA-v1}
\articletype{Article Type}%

\received{7 October 2021}
\revised{6 November 2021}
\accepted{8 November 2021}

\usepackage{graphicx}
\begin{document}

\title{Photoreverberation mapping of quasars in the context of LSST observing strategies}

\author[1]{Isidora Jankov*}

\author[1,2]{An\dj elka B. Kova{\v c}evi{\'c}}

\author[1,3]{Dragana Ili{\'c}}

\author[1,2,4]{Luka {\v C}. Popovi{\'c}}

\author[1]{Viktor Radovi{\'c}}

\author[1]{Iva {\v C}vorovi{\'c}-Hajdinjak}

\author[5]{Robert Nikutta}

\author[6]{Paula Sánchez-Sáez}

\authormark{I. Jankov \textsc{et al}}

\address[1]{\orgdiv{Department of Astronomy, Faculty of Mathematics} \orgname{University of Belgrade}, \orgaddress{\street{Studentski trg 16, 11000 Belgrade}, \country{Serbia}}}

\address[2]{\orgdiv{PIFI Research Fellow, Key Laboratory for Particle Astrophysics, Institute of High Energy Physics}, \orgname{Chinese Academy of Sciences},\orgaddress{\street{19B Yuquan Road, 100049 Beijing}, \country{China}}}

\address[3]{\orgdiv{Hamburger Sternwarte}, \orgname{Universitat Hamburg}, \orgaddress{\street{Gojenbergsweg 112, 21029 Hamburg}, \country{Germany}}}

\address[4]{\orgdiv{Astronomical Observatory},\orgname{University of Belgrade}, \orgaddress{\street{Volgina 7, 11000 Belgrade}, \country{Serbia}}}

\address[5]{\orgdiv{NSF's NOIRLab},\orgname{National Optical Astronomy Observatory}, \orgaddress{\street{950 N. Cherry Ave.,
Tucson, AZ 85719}, \country{USA}}}

\address[6]{\orgdiv{Directorate for Science},\orgname{European Southern Observatory}, \orgaddress{\street{Karl-Schwarzschild-Strasse 2, Garching bei München}, \country{Germany}}}

\corres{* \email{isidora\_jankov@matf.bg.ac.rs}}


\abstract{The upcoming photometric surveys, such as the Rubin Observatory's Legacy Survey of Space and Time (LSST) will monitor unprecedented number of active galactic nuclei (AGN) in a decade long campaign. Motivated by the science goals of LSST,  which includes the harnessing of broadband light curves of AGN for photometric reverberation mapping (PhotoRM), we implement the existing formalism to estimate the lagged response of the emission line flux to the continuum variability using only mutli-band photometric light curves. We test the PhotoRM method on a set of 19 artificial light curves simulated using a stochastic model based on the Damped Random Walk process. These light curves are sampled using different observing strategies, including the two proposed by the LSST, in order to compare the accuracy of time-lag retrieval based on different observing cadences. Additionally, we apply the same procedure for time-lag retrieval to the observed photometric light curves of NGC 4395, and compare our results to the existing literature.}



\jnlcitation{\cname{%
\author{I. Jankov}, 
\author{A. B. Kova{\v c}evi{\'c}}, 
\author{D. Ili{\'c}}, 
\author{L. {\v C}. Popovi{\'c}},
\author{V. Radovi{\'c}},
\author{I. {\v C}vorovi{\'c}-Hajdinjak},
\author{R. Nikutta} and 
\author{P. Sánchez-Sáez}} (\cyear{2021}), 
\ctitle{Photoreverberation mapping of quasars in the context of LSST observing strategies}}

\maketitle


\section{Introduction}\label{sec1}

Active galactic nuclei (AGN) are located at the center of some galaxies and represent immediate surroundings of the supermassive black hole (SMBH). They exhibit fairly complex structure, including emitting regions of different density, scale and kinematics. These are typically difficult to resolve spatially using  widely accessible observational technology in the optical band, except for the state-of-the-art interferometry {\citep[e.g.,][]{Gravity2018}}. However, some degree of spatial resolution can be obtained by including the temporal dimension in their study, which is now becoming widely available with the rapid development of time-domain astronomy (see e.g., Ili\'c et al., \citeyear{Ilic2021}, and references therein). This is the core of the reverberation mapping (RM) methods where the delayed response (i.e., time-lag) of different line emitting regions to the continuum variability occurring in the accretion disc is used to determine the radial distance of this emitting region from the central source. Since the broad-line region (BLR) kinematics are heavily influenced by the central black hole, knowing its radial distance from the central source is an important step for direct measurements of the SMBH mass.

\begin{figure*}
	\centering
	\includegraphics[width=0.47\textwidth]{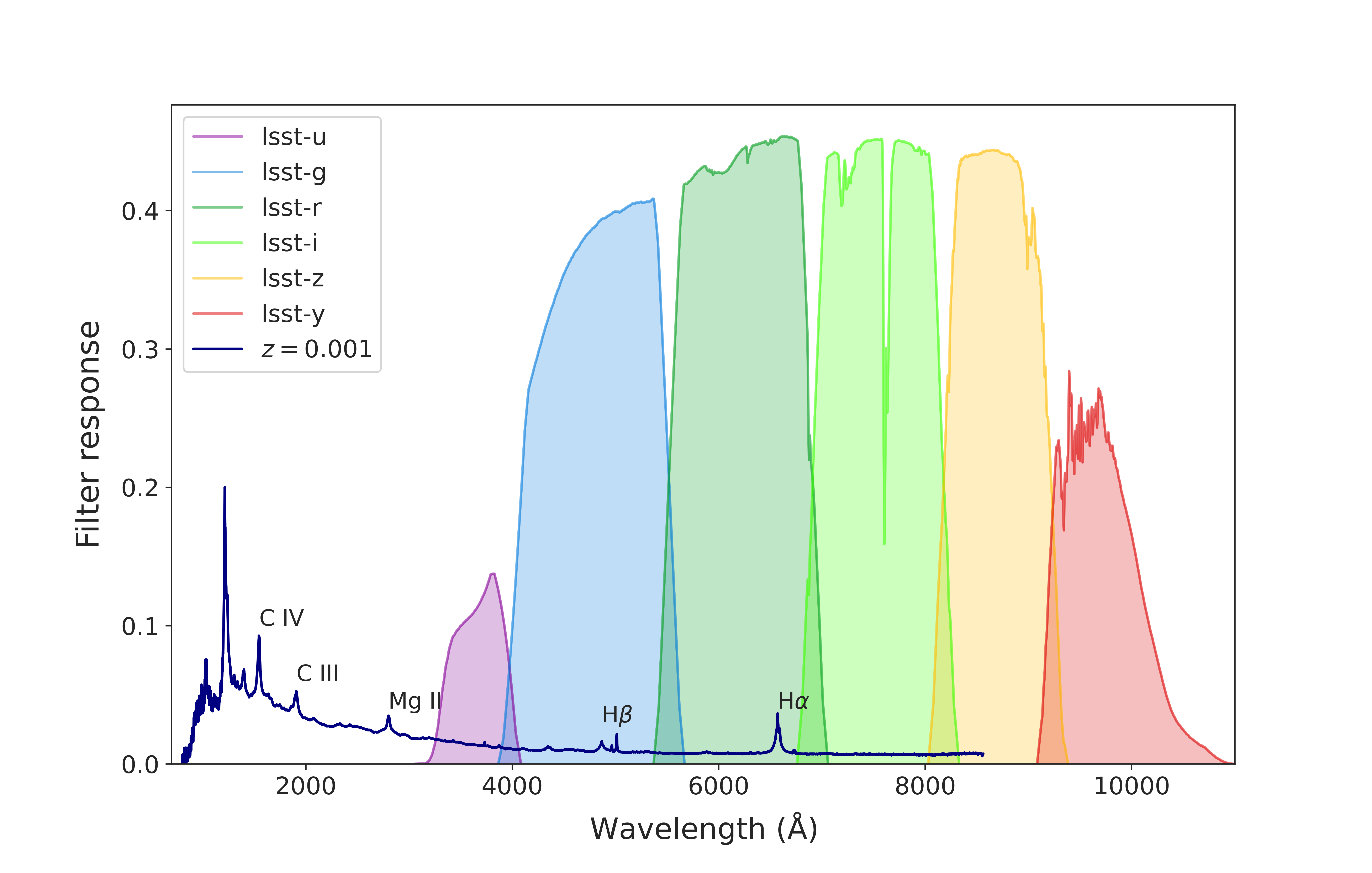}
	\includegraphics[width=0.47\textwidth]{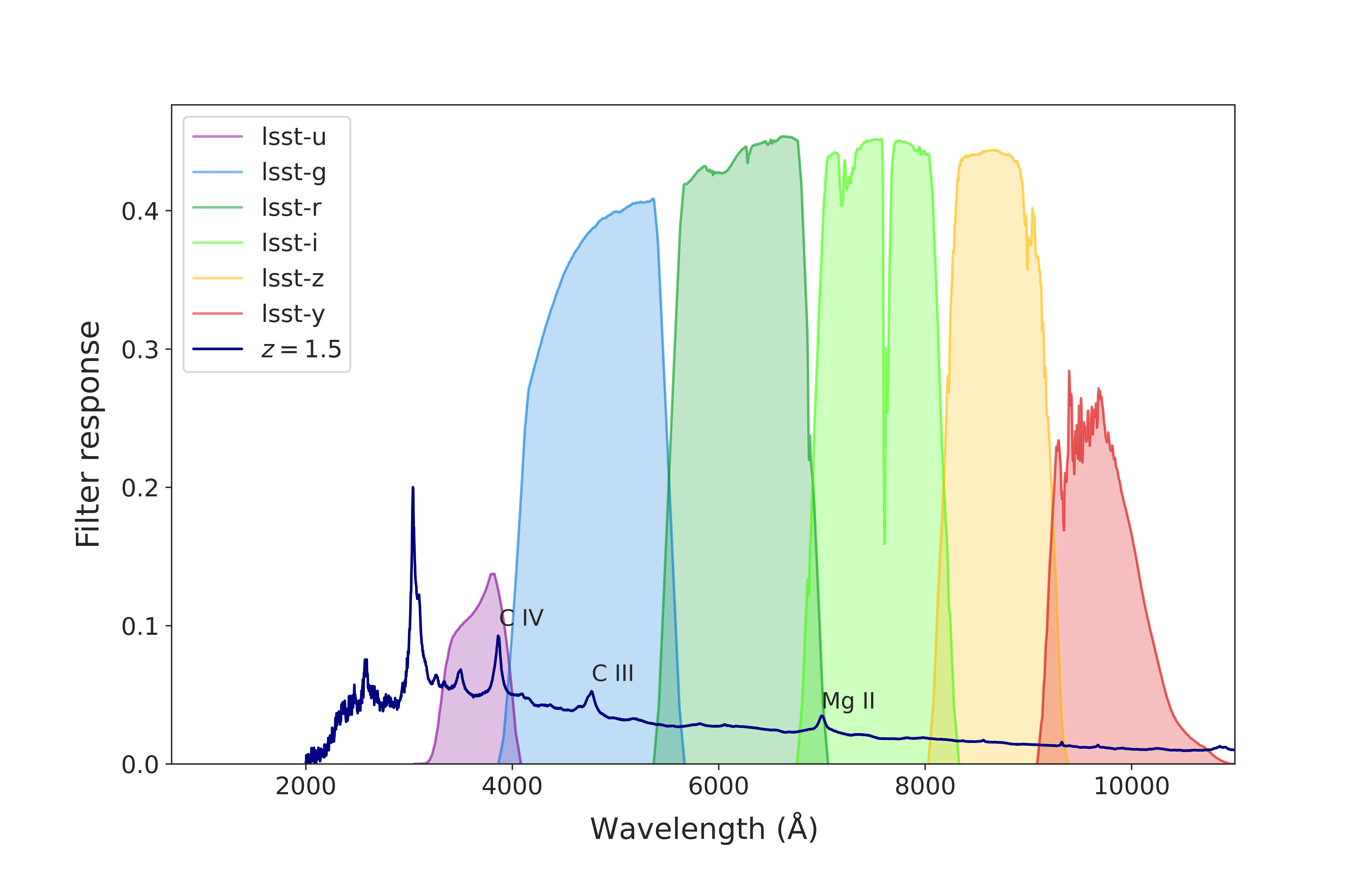}
	\caption{Composite quasar spectrum \citep{VandenBerk2001} plotted against the LSST broadband filters response curves at two different redshifts. \textit{Left:} Composite quasar spectrum with redshift of NGC 4395, $z = 0.001$. The H$\alpha$ and H$\beta$ emission lines are covered by the $r$ and $g$ bands, respectively. The pure continuum emission can be sampled via $i$ band. \textit{Right:} Composite quasar spectrum at redshift $z = 1.5$. Balmer lines are no longer visible in the spectral range covered by the LSST filters, while Mg II, C III and C IV emission lines become available for photoRM at this redshift.}  
	\label{fig1}%
\end{figure*}

The classic RM methods include careful spectroscopic monitoring of the continuum and emission line flux, which allows highly precise measurements of the BLR radius, but can be observationally time-consuming. Contrary to spectroscopic observations, photometric observations are much more time-efficient and abundant in the current astronomical databases, but also in plans for future big sky surveys. For example, the upcoming large photometric survey, the Rubin Observatory's Legacy Survey of Space and Time (LSST), is expected to monitor at least $10^7$ AGNs over the course of 10 years, providing the astronomical community an unprecedented amount of AGN light curve data. Such large quantities of data could be utilized for time-lag estimation using an efficient approach to photometric RM (PhotoRM) developed by \cite{Chelouche2012}. This method could contribute in the dramatic increase in number of reverberation mapped objects, especially in the case of faint objects which are difficult to monitor spectroscopically, and they will be highly present in the LSST Data Releases.

The main goal of this paper is to asses the accuracy of time-lag estimation of our Python implementation of PhotoRM method based on the formalism by \cite{Chelouche2012} by performing several tests on the multi-band photometric light curves with different cadence strategies. The tests address the baseline accuracy of this method using the artificially generated light curves sampled using 1 day cadence, but also using several different non-ideal observing strategies, including the light curves with hypothetical variable cadence, cadences from LSST Operation Simulator \citep[OpSim,][]{Jones2020} and real observational light curve of NGC 4395 (Edri et al., \citeyear{Edri2012}).

The light curve simulation process, observing strategies used and the description of the observational data is given in Section \ref{sec2}. In Section \ref{sec3}, time-lags are measured using our Python implementation of PhotoRM where we apply it to previously mentioned observing strategies and observational data of NGC 4395. In the case of simulated data, obtained values are compared to true values used for light curve simulation, whilst for NGC 4395, we compared measured time-lags to ones from the available literature. Our conclusions are presented in Section \ref{sec4}.

\begin{figure}[ht]
	\centerline{\includegraphics[width=90mm,height=85mm]{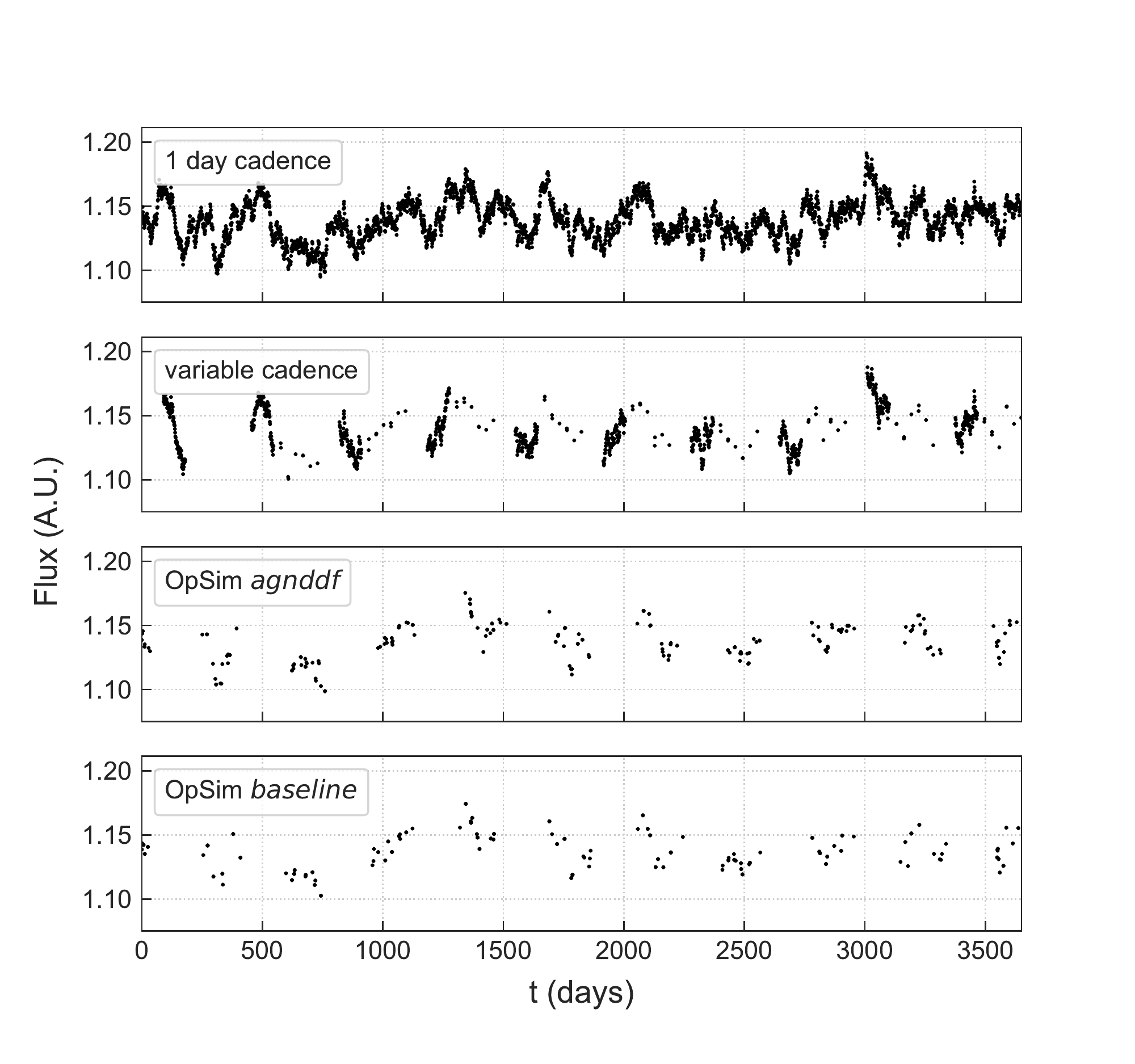}}
	\caption{An example of the simulated light curves in an arbitrary broadband filter (ID=13, see Table \ref{tab1}) sampled using different observing strategies. From top to bottom: ideal observing strategy (1 day cadence); hypothetical variable cadence; cadence from OpSim run \textit{agnddf\_v1.5\_10yrs}; cadence from OpSim run \textit{baseline\_samefilt\_v1.5\_10yrs}. All observing strategies have the same length of 10 years.}
	\label{fig2}
\end{figure}

\begin{figure*}
	\centerline{\includegraphics[width=184mm,height=85mm]{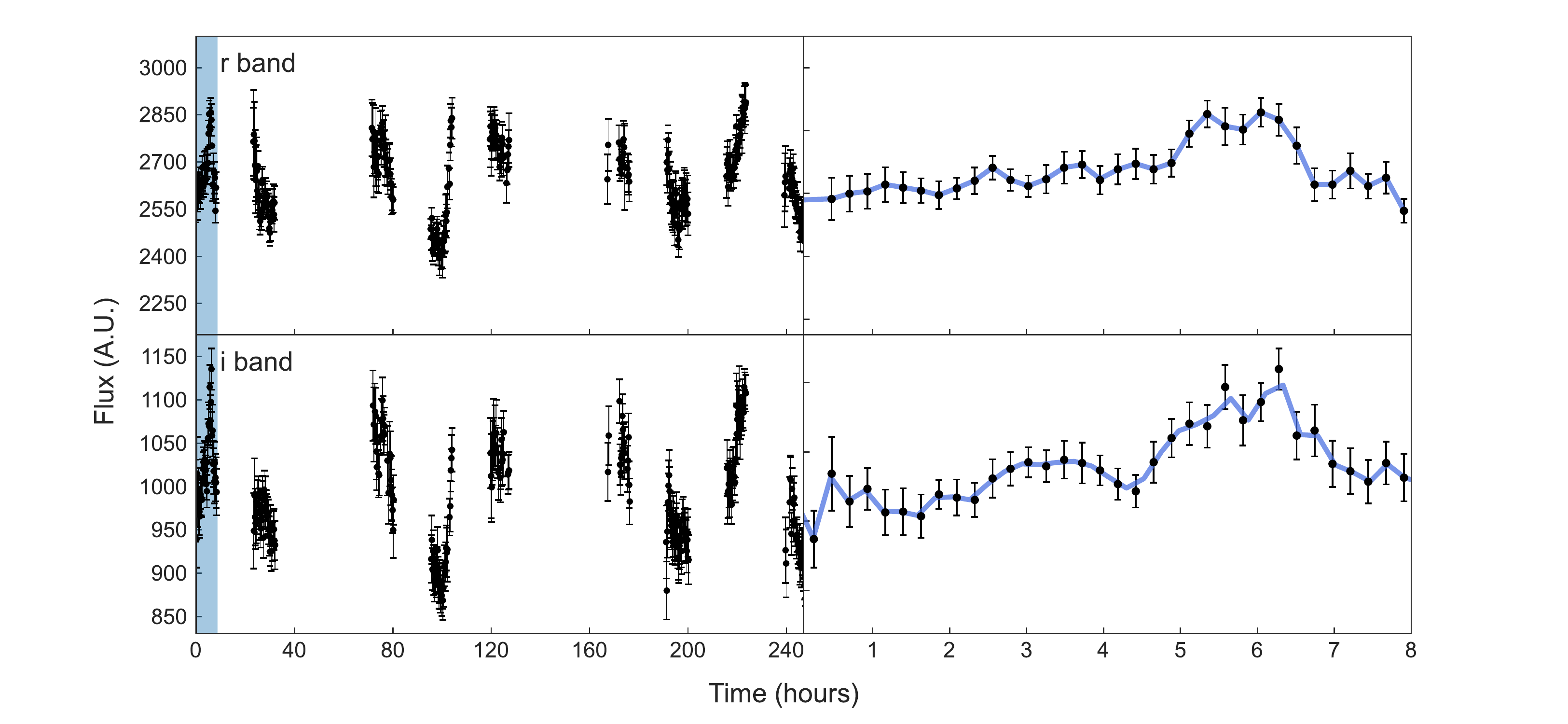}}
	\caption{Observed light curves of NGC 4395 in $r$ (upper panels) and $i$ bands (bottom panels) from Edri et al., (\citeyear{Edri2012}). \textit{Left}: Light curves during the total monitoring period of 9 nights. Shaded areas represent the data range chosen for further analysis (first $\sim$8h). \textit{Right}: Data from the first 8 hours of observations ($\sim$35 points). Light blue line is the model light curve obtained using CNP.}
	\label{fig3}
\end{figure*}

\begin{figure}
	\centering
	\includegraphics[width=0.43\textwidth]{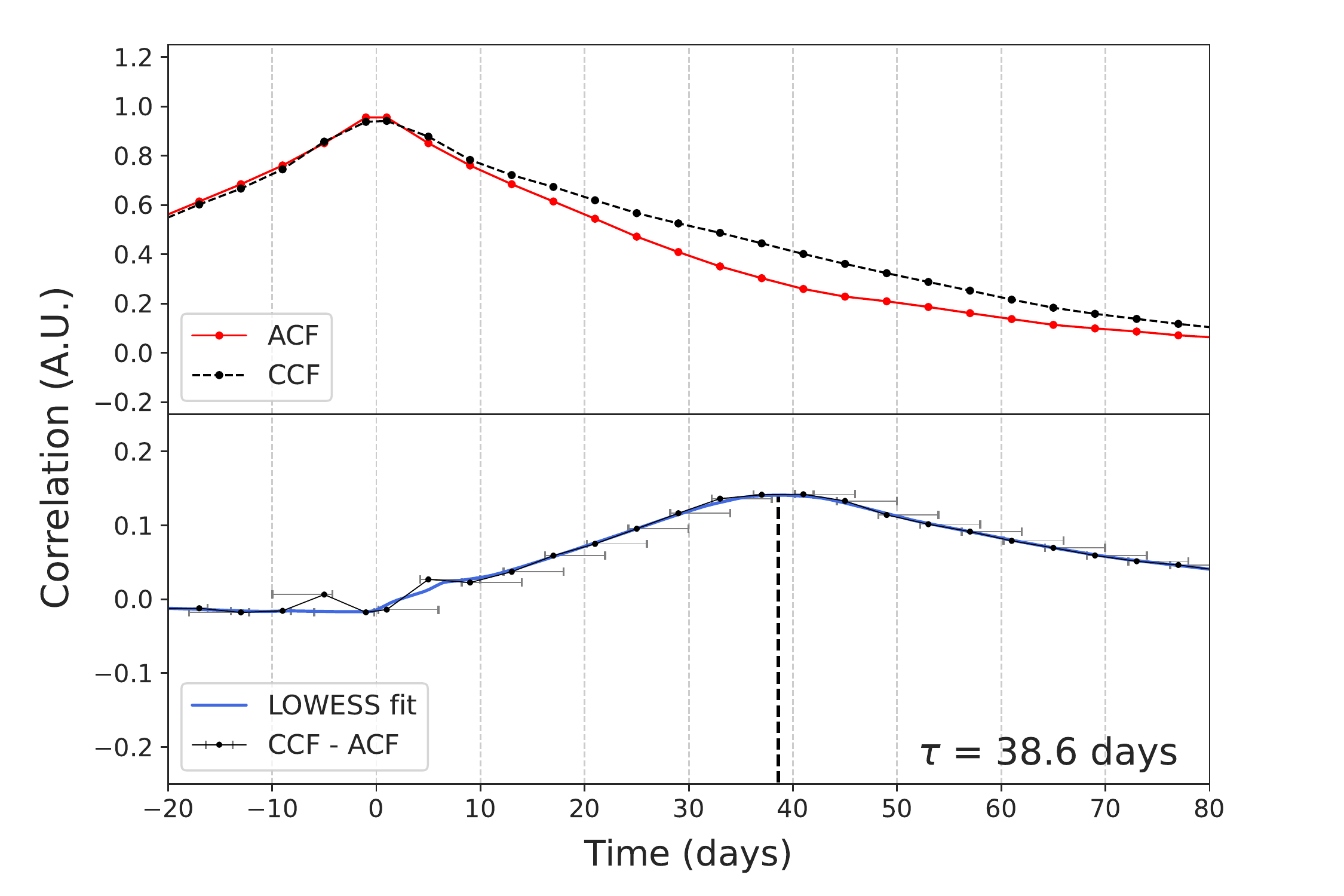}
	\includegraphics[width=0.43\textwidth]{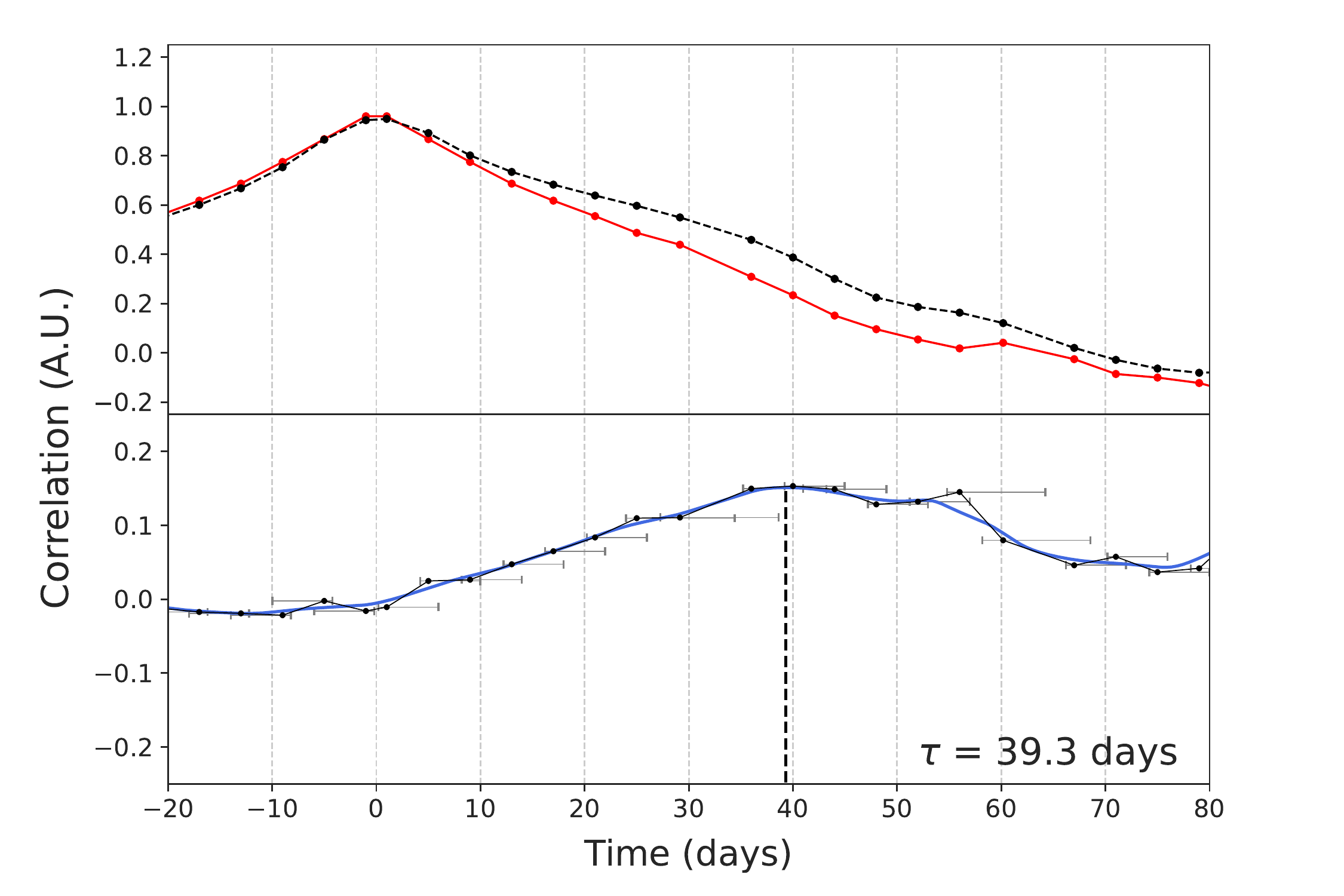}
	\includegraphics[width=0.43\textwidth]{figures/Figure_4_top.pdf}
	\caption{CCF (dashed), ACF (solid) and their difference (lower panels) for simulated AGN light curves with different cadences (Table \ref{tab1}, Object ID=13). \textit{Top}: ideal  1 day cadence; \textit{Middle}: the hypothetical variable cadence; \textit{Bottom}: the "agnddf" LSST OpSim cadence. Second order spline interpolation and the LOWESS fit (blue solid line) are used together to obtain the difference between the CCF and ACF. True value of the time-lag is $\tau_{\mathrm{true}}=36.9$ days, and estimated ones are indicated on each panel in the bottom right.}
	\label{fig4}%
\end{figure}

\section{Method and Data}\label{sec2}

Taking into account that different broadband filters cover different groups of spectral features, it is expected that a broadband photometric light curve will not be able to trace emission line variability separately from the continuum. Fig. \ref{fig1} illustrates this by plotting composite quasar spectrum against LSST filter response curves, clearly revealing that the emission lines are entangled with the continuum emission from the broadband filter perspective (e.g., H$\alpha$ line and its surrounding continuum in the $r$ band on the left panel of Fig. \ref{fig1}). To account for this, \cite{Chelouche2012} developed an efficient, although less sensitive, method which allows the utilization of photometric light curves for time-lag estimation. In their formalism, they select light curves in two carefully chosen filters (for an example of filter choice at different redshifts see Fig. \ref{fig1}), where one of those must cover the pure continuum emission and the other covers an emission line and the surrounding continuum (X and Y bands, respectively, further in the text). To successfully disentangle the continuum and line emission, the line which will be covered must be chosen in accordance to the broadband filters coverage at given redshift (Fig. \ref{fig1}) and have significant contribution in the total integrated flux. The cross-correlation function between the line emission hidden in the broadband light curve and pure continuum emission is obtained using the time-lag estimator which is based on the difference between cross-correlation (CCF) and auto-correlation functions (ACF) (Edri et al. \citeyear{Edri2012}, their Eq. 4):
\begin{align}
    \label{eq1}
    CCF(\tau)\approx CCF_{XY}(\tau) - ACF_X(\tau)
\end{align}
The method assumes that the time variability of the continuum flux in the X band is the same as that in the Y band, which is a fair approximation since the continuum contributes 75-95\% of the total flux in the filters, while the remaining flux comes from the broad emission lines (Edri et al., \citeyear{Edri2012}). The time-lag is estimated using the obtained line-continuum CCF. The $z$-transformed discrete correlation function (ZDCF) by \cite{Alexander1997} was used for CCF calculation, since it can deal with light curves having distribution of points as considered LSST OpSim cadence and artificial cadence.

In order to test our implementation of the photoRM method, we perform several experiments with a set of 19 artificially generated pairs of light curves of the continuum and line emission with different time-lags ranging from a few days to several months. The light curves were generated as described in \cite{Kovacevic2021b}, and are further entangled using the prescription in \cite{Chelouche2012}. The procedure involves the entanglement of the generated continuum and emission line light curves (Eq. 4. Chelouche \& Daniel, \citeyear{Chelouche2012}) in  order to simulate a kind of light curve we get from the broadband photometric filters. 

We shortly describe how the artificial light curves are simulated, for more details see \cite{Kovacevic2021b}. The continuum light curves were generated using the Damped Random Walk (DRW) model (Kelly et al. \citeyear{Kelly2009}), with the characteristic amplitude $\sigma$ and timescale $\tilde{\tau}$ inferred from the SMBH mass and/or AGN luminosity. The DRW model parameters, $\sigma$ and $\tilde{\tau}$, are calculated from  the  first principles: assuming \textit{a prirori} theoretical distributions of AGN luminosity, SMBH mass and BLR radius as given in \cite{Kovacevic2021b}.
The simulated light curves have idealized cadence of 1 day over the interval of 5000 days (13.7 years). The chosen interval satisfies the condition that the length of the light curve used for time-lag estimation must be at least 10 times larger than its characteristic timescale $\tilde{\tau}$ \citep{Kozlowski2017}. From there, the emission line light curves were obtained using the linear approximation:
\begin{align} 
  \label{eqn2}
    f^{l} (t) =  (f^{c}*\xi)(t)
\end{align}
where the transfer function $\xi$ defines the geometry of the BLR as seen by the observer and $f^c$ is the continuum flux obtained using the DRW model, as described above. For simplicity, $\xi$ is approximated by the Gaussian kernel with the mean equal to the BLR radius $R_{\mathrm{BLR}}$ and a standard deviation set to $0.25 R_{\mathrm{BLR}}$ \citep{Chelouche2012}. The $R_{\mathrm{BLR}}$ is calculated from the SMBH mass used for generating DRW model parameters. When calculating the entangled broadband light curves, which contain the contribution from both continuum and line emission, we assume in all cases that the emission line contributes with 20\% to the total broadband flux. This level of contribution is moderately increased compared to the one for H$\alpha$ emission in real light curves {\citep[e.g., $\eta = 7\%$ for NGC 4395,][]{Edri2012}} in order to mitigate the problem of large uncertainties in extraction of time-lags from light curves with such a small emission line contributions (Zu et al., \citeyear{Zu2016}). In principle, theoretical framework behind Equation \ref{eq1} allows for a wider range of values (e.g., $\eta =$ 10-50\%) to be assumed as long as the emission line contribution is sub-dominant to the continuum contribution \citep{Chelouche2012}. Next, we constructed the set of light curve pairs representing the broadband photometric light curves, one containing only the continuum emission and the other the emission both from the continuum and line sources, for which the time-lag between the continuum and line emission is {\it a priori} known. Before testing the extraction of the time-lag, all simulated light curves were cut to a 10 year baseline in order to mimic the length of LSST operation period. 
The accuracy of time-lag retrieval from these simulated broadband photometric light curves is evaluated for several different observing strategies which are visually demonstrated on one example of the simulated light curve (Fig. \ref{fig2}). The considered observing strategies include:

\begin{enumerate}[1.]
\item"Ideal" observing strategy: light curves sampled with cadence of 1 day for the whole duration of the 10 year survey period.

\item"Variable" observing strategy: the first year has 3 months of observations with 1 day cadence, and in the remaining years of the 10 year period the observations are carried out with 1 day cadence during the first 3 months, followed by 6 months of 30-day cadence and a gap of 3 months. This observing strategy is constructed using the criteria involving the number of observations and inclusion of alternating seasons (years).

\item Two examples of LSST OpSim runs used in our LSST AGN Science Collaboration Cadence Note \citep{Kovacevic2021a}. The first one is \textit{agnddf\_v1.5\_10yrs\_r\_ra\_0.0\_de\_-0.52} and it is designed with AGNs in mind ("agnddf", further in the text). This OpSim run involves the deep drilling fields which represent several fields in the southern sky with larger number of visits compared to the rest of the footprint. The other OpSim run is the \textit{baseline\_samefilt\_v1.5\_10yrs\_r\_ra\_0.0\_de\_-0.52} ("baseline", further in the text) observing strategy which was chosen as a reference point for comparison since it involves the standard LSST footprint. For the purpose of this study, both OpSim cadence strategies are taken from the $r$-band. The OpSim outputs were obtained from Feature Based Scheduler release 1.5\footnote{http://astro-lsst-01.astro.washington.edu:8081/} (FBS 1.5).

\end{enumerate}

\begin{figure*}[ht]
	\centering
	\includegraphics[width=0.47\textwidth]{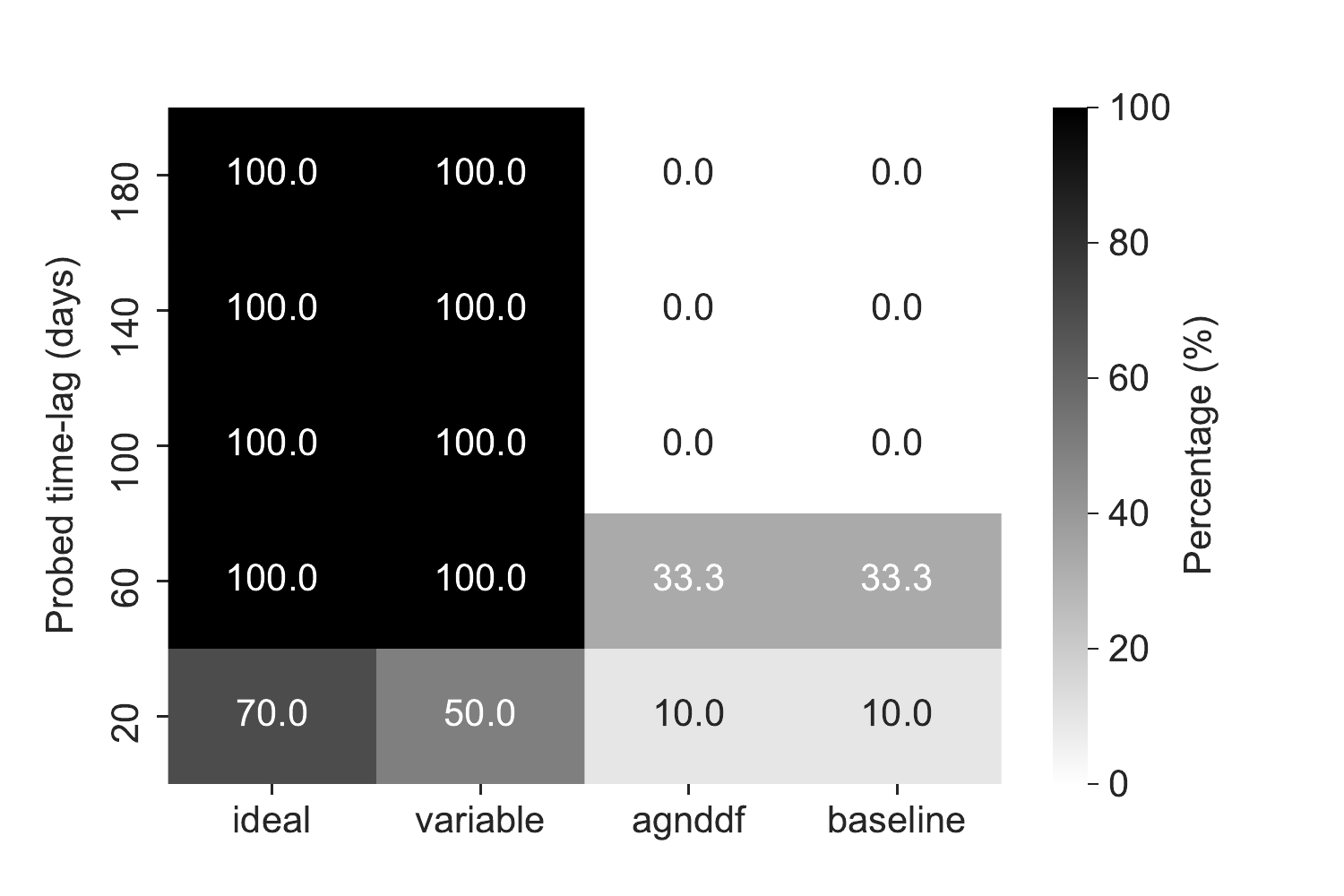}
	\includegraphics[width=0.47\textwidth]{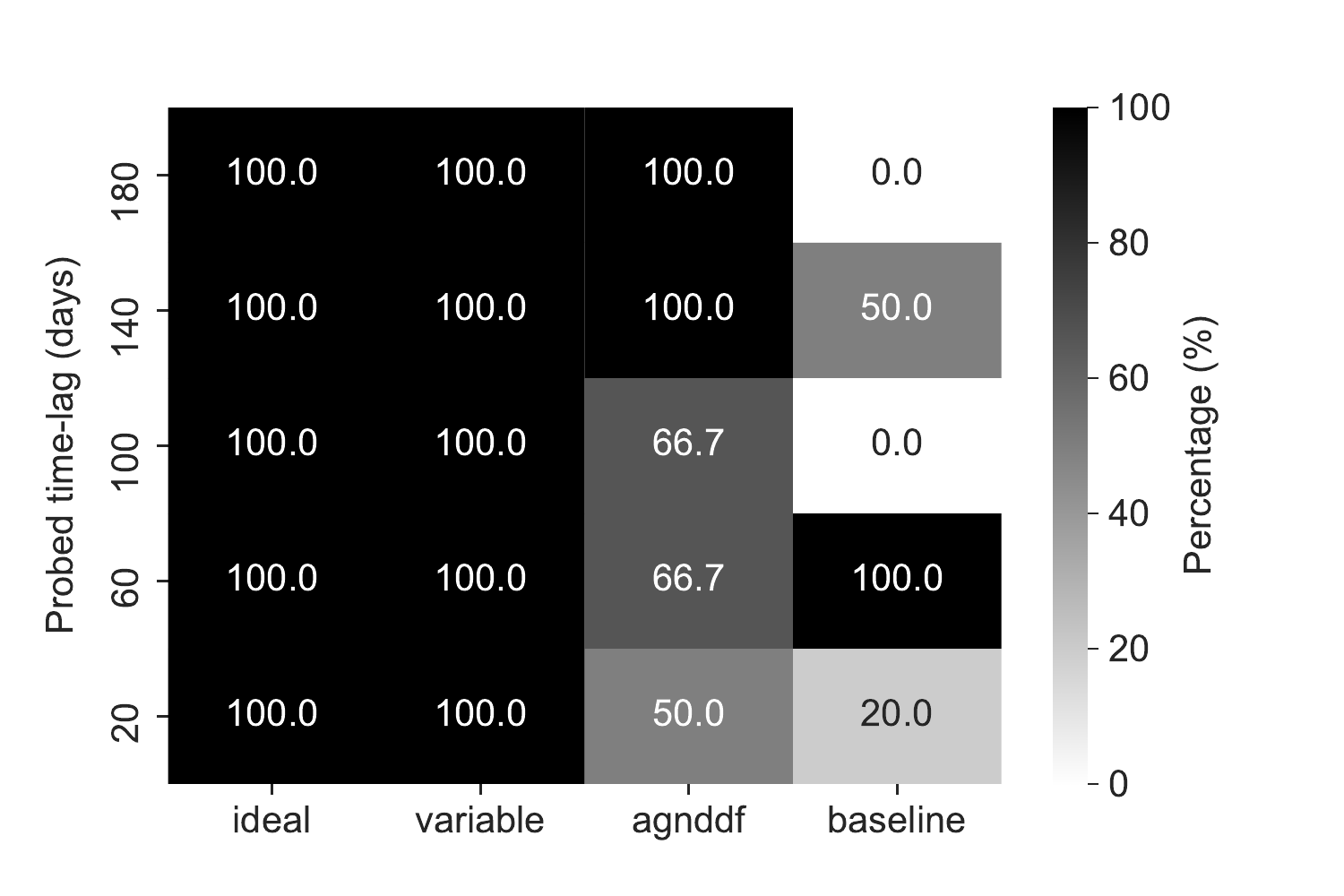}
	\caption{Time-lag detection efficiency for different observing strategies. Numbers indicate the percentage of estimations per time-lag bin with relative error bellow 10\% (left) and bellow 30\% (right).}  
	\label{fig5}%
\end{figure*}

In addition to the simulated light curves, we test our implementation of photoRM method on the observed photometric light curves of NGC 4395 obtained during the 9-night monitoring period conducted by \cite{Edri2012}. Observations were performed by Wise Observatory's 1m telescope using the Sloan Digital Sky Survey's $g$, $r$ and $i$ bands. In this case study, we have used only the $r$ and $i$ bands (Fig. \ref{fig3}), since the $r$-band covers the H$\alpha$ line which has larger contribution to the total flux than the H$\beta$ line covered by the $g$-band. The $i$-band covers the pure continuum emission. Because NGC 4395 is a low luminous Seyfert 1 galaxy, the expected time-lag is only of the order of a few hours, as confirmed by the spectroscopic RM campaigns (Peterson et al., \citeyear{Peterson2005}, Desroches et al., \citeyear{Desroches2006}). This allows us to avoid the gaps in the light curve by cutting it so it only contains the data from the first night, i.e. first $\sim8$ hours. For comparison, we also apply the Conditional Neural Process (CNP, Garnelo et al., \citeyear{Garnelo2018}) to the cut light curves in order to test whether time-lag retrieval accuracy is improved (see Fig. \ref{fig3}, right panels). Our formalism describing the CNP light curves is presented in details in \v{C}vorovi\'{c} - Hajdinjak et al. within this special issue.

\section{Results and Discussion}\label{sec3}

Using the \cite{Chelouche2012} approach to photoRM and ZDCF for calculation of CCFs we obtain time-lags of the emission line flux response to the continuum flux variation from 19 simulated pairs of broadband light curves. Each set of light curves has four time-lag estimations, each corresponding to different observing strategy (Table \ref{tab1}). To obtain time-lags, we developed a Python code which implements the previously described procedure for the line-continuum CCF calculation, but also for locating the first prominent peak in the obtained CCF which corresponds to the actual time-lag. The time-lag errors were calculated using the asymmetric error propagation method described in Laursen et al. (\citeyear{Laursen2019}). The example analysis of one of the simulated light curves using different observing strategies is shown in Fig. \ref{fig4}. In many cases, especially for light curves with smaller and fewer gaps (i.e., in ideal and variable strategies), we have found that the prediction accuracy is improved if the CCF is fitted using the locally weighted scatterplot smoothing\footnote{Code by Alexandre Gramfort: https://gist.github.com/agramfort/850437} (LOWESS, Cleveland \& Devlin, \citeyear{Cleveland1988}), a form of non-parametric regression. Although CCFs in our analysis do not have large scatter, this method is still useful for smoothing sharp drops and peaks which sometimes appear in CCF-ACF difference. We observed that the prediction accuracy drops for $\tau < 20$, because the peaks tend to be defined by a single outlying point, therefore we did not use the fitting method in those cases. As demonstrated in Fig. \ref{fig4}, very accurate prediction can be obtained using this method, although prediction accuracy goes down as the number of points contained in the light curve decreases (e.g., comparing ideal and variable cadence where number of points drops by a factor of $\sim$4) but also depends on the choice of the observing cadence. For example, OpSim "agnddf" and "baseline" observing strategies have similar number of points ($\sim200$), as well as the mean cadence (16.8 and 17.9 days, respectively), but vary in time-lag retrieval accuracy. This most likely occurs due to less prominent features in light curves sampled using the "baseline" cadence (see bottom  panel in Fig. \ref{fig2}), making it difficult to account for peaks and valleys needed in order to detect the time-lag in CCF-ACF difference.

The time-lag detection efficiency is shown in Fig. \ref{fig5}, indicating the percentage of time-lag estimations belonging to different time-lag ranges and observing strategies, with relative errors bellow 10\% (left panel) and 30\% (right panel). As expected, continuous light curves tend to give the best predictions for 10\% error margin, with highest detection efficiency for $\tau > 40$ days and with total mean squared error (MSE) of 3.9. The hypothetical variable cadence has comparable time-lag detection efficiency and $\mathrm{MSE_{var}}$ of only 10.9. Of two OpSim realization, the "agnddf" performed better when considering detection efficiency with 30\% error margin, whereas at the 10\% error it didn't show better performance than "baseline" cadence. Right panel of Fig. \ref{fig5} also shows that "agnddf" performs better in detection of longer time-lags, whilst the "baseline" is only good for $40 < \tau < 80$.  MSE scores also indicate that "agnddf" has better detection efficiency ($\mathrm{MSE_{agnddf}} = 377.2$) compared to "baseline" ($\mathrm{MSE_{baseline}} = 1305.6$). As input time lags of each artificial object are inferred from the theoretical distribution given in \cite{Kelly2009}, larger set of light curves will be tested in the forthcoming paper.

\begin{figure*}[ht]
	\centering
	\includegraphics[width=0.47\textwidth]{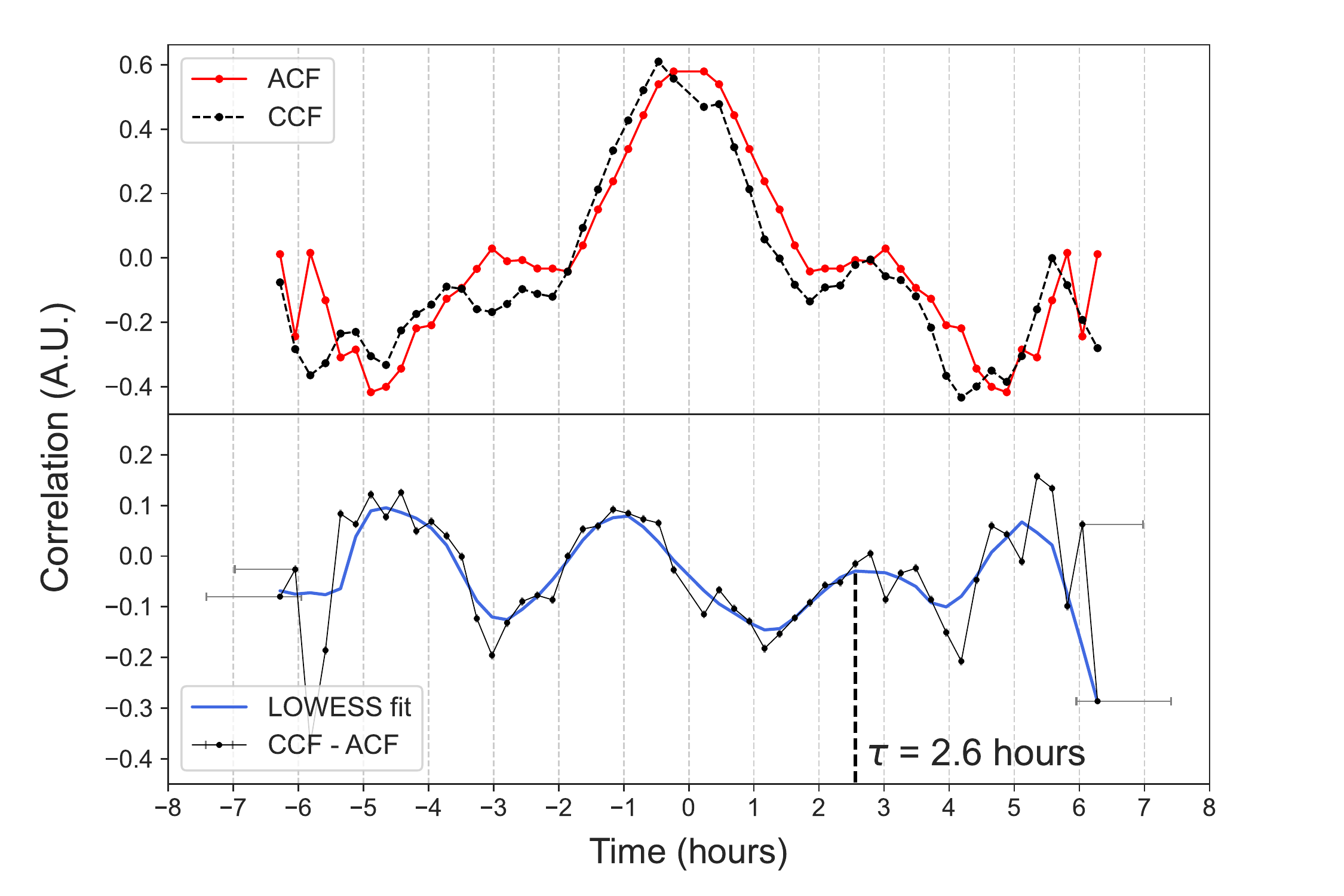}
	\includegraphics[width=0.47\textwidth]{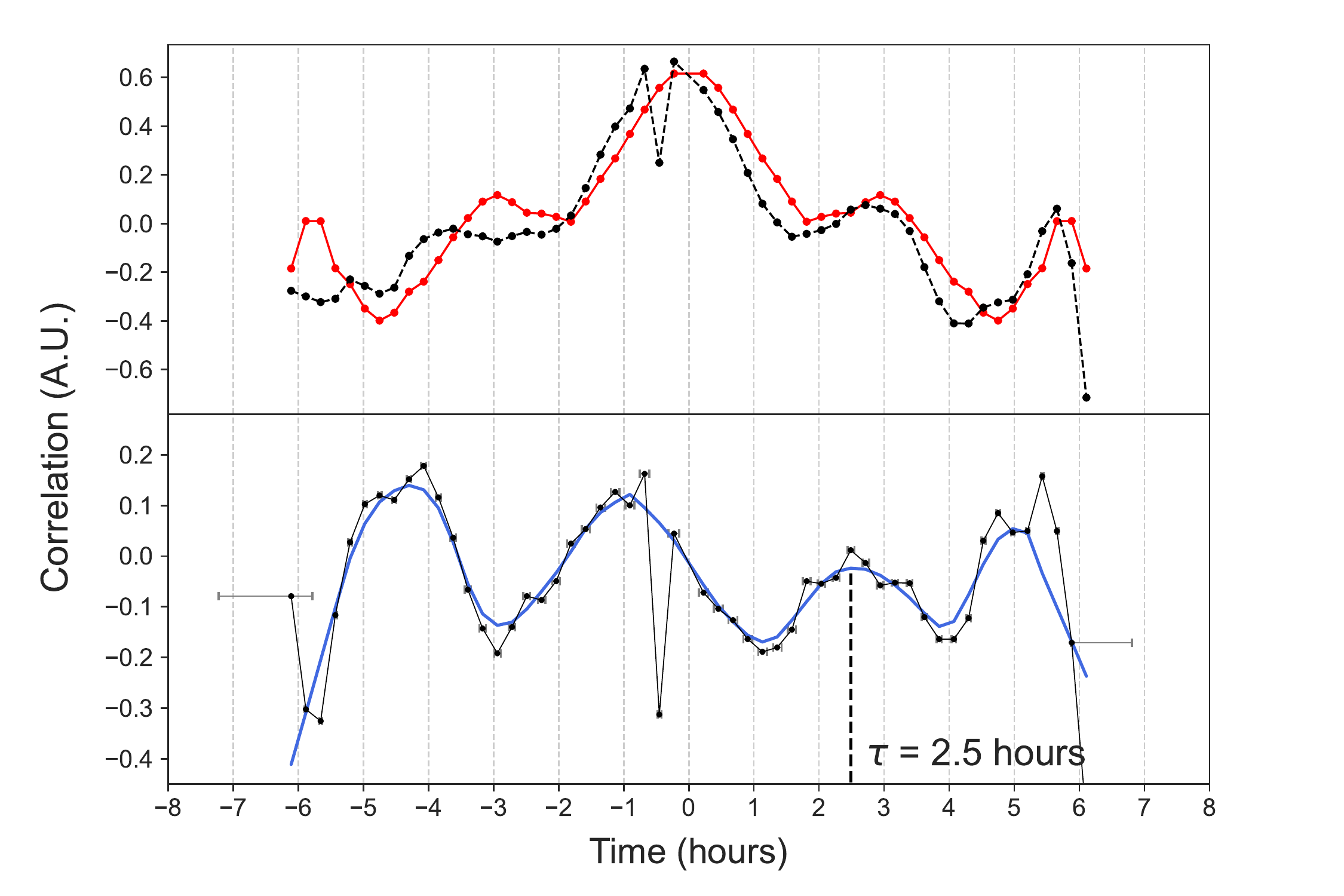}
	\caption{Cross-correlation analysis and time-lag estimation for NGC 4395 for the 8 hour long observed light curve segment (left) and the CNP modeled light curve segment (right). The notation is the same as in Fig. \ref{fig4}. For comparison see \cite{Edri2012} and their Fig. 8.}  
	\label{fig6}%
\end{figure*}

The second aspect of our analysis is utilization of the same procedure, but now applied to the observed light curve of NGC 4395 (Fig. \ref{fig3}), a well-known low luminosity AGN which was a subject of several RM campaigns \citep[][]{Peterson2005, Desroches2006,Edri2012}. Since the $r$-band covers the H$\alpha$ emission line and $i$-band covers the pure continuum, we cross-correlate the $i$ and $r$ band, similarly as in \cite{Edri2012}.  We found that the H$\alpha$ emission lags $2.6^{+0.002}_{-0.001}$ hours behind the continuum emission using only 35 data points contained in the analyzed light curve segment (left panel in Fig. \ref{fig6}). When the same light curve segment is approximated by the CNP model, we estimated comparable time-lag value of $2.5^{+0.06}_{-0.05}$ hours (right panel in Fig. \ref{fig6}). Both extracted values for time-lag are comparable to the results of previous photoRM \citep[$\tau = 3.6$ hours,][]{Edri2012} and spectrsocopic RM estimates for H$\alpha$ time-lag \cite[$\tau = 1.44$ hours,][]{Desroches2006}.

\begin{center}
\begin{table*}[t]
\caption{Summary of simulated light curve properties and measured time-lags and their errors for different observing strategies. The columns are: (1) light curve ID, (2) bolometric luminosity, (3) SMBH mass, (4) BLR dimension, (5) time-lag for "ideal" cadence (3650 points), (6) time-lag for "variable" cadence (1000 points), (7) time-lag for LSST OpSim "agnddf" cadence (217 points), (8) time-lag for LSST OpSim "baseline" cadence (204 points). All simulated light curves are 10 years long. The last row displays measured H$\alpha$ time-lags using photometric light curves of NGC 4395:
$\tau_{\mathrm{obs}}$ is obtained from the 8 hour long light curve segment and  $\tau_{\mathrm{CNP}}$ is measured from the CNP model of the same 8 hour long segment. \label{tab1}}
\centering
\begin{tabular*}{500pt}{@{\extracolsep\fill}@{}llllllll@{}}
\toprule
&\multicolumn{3}{@{}c@{}}{\textbf{Input parameters for DRW}}&\multicolumn{2}{@{}c@{}}{\textbf{Non-LSST cadence}} & \multicolumn{2}{@{}c@{}}{\textbf{LSST cadence}} \\\cmidrule{2-4}\cmidrule{5-6}\cmidrule{7-8}
\textbf{ID} 
& \textbf{$\mathrm{log \ }L_{\mathrm{bol}}$} 
& \textbf{$M_{\mathrm{BH}}$} 
& \textbf{$R_{\mathrm{BLR}}$} 
& \textbf{$\tau_{\mathrm{1d}}$}
& \textbf{$\tau_{\mathrm{var}}$}
& \textbf{$\tau_{\mathrm{agnddf}}$}
& \textbf{$\tau_{\mathrm{baseline}}$}
\cr 
& 
& $\mathrm{[10^6} \ M_{\mathrm{\odot}}]$ 
& $[\mathrm{1d}]$
& $[\mathrm{1d}]$ 
& $[\mathrm{1d}]$ 
& $[\mathrm{1d}]$ 
& $[\mathrm{1d}]$ 
\cr (1) & (2) & (3) & (4) & (5) & (6) & (7) & (8)
\\ \midrule
1  &  44.01    & 14.82   & 34.1   & $36^{+5}_{-0.8}$   & $43^{+5}_{-0.8}$        & $42^{+17.9}_{-12}$    & $47.9^{+19.3}_{-11.6}$      \\[0.3cm]
2   & 44.56    & 34.42   & 66.9   & $67.6^{+5}_{-0.8}$     & $64.8^{+5}_{-0.8}$      & $43.1^{+17.9}_{-12}$  & $51.4^{+19.3}_{-11.6}$      \\[0.3cm]
3   & 44.19    & 19.47   & 42.4   & $41.6^{+5}_{-0.8}$   & $39.3^{+5}_{-0.8}$      & $43.3^{+17.9}_{-12}$  & $50.2^{+19.3}_{-11.6}$      \\[0.3cm]
4   & 44.00    & 14.52   & 33.6   & $34.5^{+5}_{-0.8}$   & $40.9^{+5}_{-0.8}$      & $45.6^{+17.9}_{-12}$  & $51.5^{+19.3}_{-11.6}$      \\[0.3cm]
5   & 43.60    & 7.86    & 20.5   & $25^{+5}_{-0.8}$   & $25^{+9.4}_{-1}$        & $20.1^{+24.3}_{-12.5}$  & $44^{+19.3}_{-11.6}$      \\[0.3cm]
6   & 44.33    & 24.23   & 50.5   & $50.5^{+5}_{-0.8}$   & $47.8^{+5}_{-0.8}$      & $39.1^{+17.9}_{-12}$    & $50.3^{+19.3}_{-11.6}$      \\[0.3cm]
7   & 43.64    & 8.43    & 21.7   & $25^{+5}_{-0.8}$   & $25^{+9.4}_{-1}$       & $27^{+17.9}_{-12}$       & $22.5^{+29.9}_{-18.7}$  \\[0.3cm]
8   & 45.14    & 84.19   & 136.8  & $134.3^{+5}_{-0.8}$  & $138.8^{+5}_{-0.8}$     & $120.8^{+8.7}_{-5.6}$    & $172.1^{+24.4}_{-11.9}$ \\[0.3cm]
9   & 45.38    & 121.18  & 183.1  & $185.4^{+5}_{-0.8}$  & $184.1^{+9}_{-2.2}$     & $146.5^{+8.8}_{-5.4}$   & $112.6^{+12.1}_{-10.8}$ \\[0.3cm]
10  & 43.46    & 6.41    & 17.5   & $17^{+5}_{-0.8}$   & $17^{+5}_{-0.8}$        & $34.6^{+17.9}_{-12}$    & $15.7^{+29.9}_{-18.7}$  \\[0.3cm]
11  & 44.81    & 50.72   & 91.2   & $95.1^{+5}_{-0.8}$   & $90.2^{+7.6}_{-2.6}$     & $68.4^{+19.5}_{-11.3}$  & $50.3^{+19.3}_{-11.6}$      \\[0.3cm]
12  & 43.94    & 13.39   & 31.5   & $31.9^{+5}_{-0.8}$   & $32.2^{+9.5}_{-1.9}$    & $39^{+17.9}_{-12}$       & $47.4^{+19.3}_{-11.6}$      \\[0.3cm]
13  & 44.07    & 16.33   & 36.9   & $38.6^{+5}_{-0.8}$   & $39.3^{+5}_{-0.8}$      & $41.7^{+17.9}_{-12}$     & $48.8^{+19.3}_{-11.6}$      \\[0.3cm]
14  & 45.25    & 99.93   & 156.9  & $157^{+5}_{-0.8}$  & $158.1^{+5}_{-0.8}$     & $120.8^{+8.7}_{-5.6}$    & $83.6^{+17.7}_{-10.9}$  \\[0.3cm]
15  & 42.43    & 1.32    & 4.9    & $5^{+5}_{-0.8}$    & $5^{+5}_{-0.8}$         & $13^{+24.3}_{-12.5}$    & $48.5^{+19.3}_{-11.6}$      \\[0.3cm]
16  & 42.49    & 1.43    & 5.3    & $5^{+5}_{-0.8}$    & $5^{+5}_{-0.8}$         & $13^{+24.3}_{-12.5}$    & $48.5^{+19.3}_{-11.6}$      \\[0.3cm]
17  & 42.27    & 1.02    & 4.0    & $5^{+5}_{-0.8}$    & $5^{+5}_{-0.8}$         & $13^{+24.3}_{-12.5}$    & $48.5^{+19.3}_{-11.6}$      \\[0.3cm]
18  & 44.95    & 62.39   & 107.7  & $110.3^{+5}_{-0.8}$  & $106.8^{+5}_{-0.8}$     & $61.9^{+19.5}_{-11.3}$  & $51.2^{+19.3}_{-11.6}$      \\[0.3cm]
19  & 44.77    & 47.35   & 86.4   & $86.7^{+5}_{-0.8}$   & $84^{+4}_{-0.8}$        & $74.5^{+19.5}_{-11.3}$  & $57.2^{+19.3}_{-11.6}$      \\[0.3cm]
\midrule
\\
NGC 4395 & $\tau_{\mathrm{obs}} \ \mathrm{[1h]} = 2.6^{+0.002}_{-0.001}$ & $\tau_{\mathrm{CNP}} \ \mathrm{[1h]} = 2.5^{+0.06}_{-0.05}$ & & & & &\\[0.4cm]
\bottomrule
\end{tabular*}
\end{table*}
\end{center}

\section{Conclusions}\label{sec4}

To test the time-lag estimation using photoRM as described in \cite{Chelouche2012}, we have implemented the method in programming language Python using the ZDCF for CCF calculations. We have tested our method on 19 pairs of 10 year-long artificial light curves, simulated using the DRW model, as well as on the 8 hour-long observed light curves of NGC 4395 from \cite{Edri2012}. The long-term light curves are sampled using several different observing strategies, ideal 1-day cadence, variable cadence, and two LSST OpSim cadences, in order to compare the accuracy of time-lag retrieval based on different observing cadences. The two LSST OpSim runs ("baseline" and "agnddf") were chosen only as a case study to test the method itself. Moreover, we applied the photoRM method to NGC 4395 light curves modeled with CNP, in difference to \cite{Edri2012} where the classical interpolation was used. From our analysis we outline the following:
\begin{enumerate}[(i)]
    \item The photoRM method can be used for the time-lag extraction from the simulated light curves for all considered observing strategies. The prediction accuracy drops for sparsely sampled light curves, but also depends on the specific kind of observing strategy used. The same is for the observed light curves of NGC 4395, for which the obtained time-lags are comparable to the previous both photometric and spectroscopic estimates, with deviations of $\sim$1 hour in both cases.
    \item Two LSST OpSim observing strategies were compared, each having approximately same number of points and mean cadence, but different time-lag retrieval accuracy due to different considerations in cadence design. Specifically, "agnddf" performed better than "baseline" observing strategy, as expected. 
\end{enumerate}
In further analysis, we plan to assess a large set of LSST cadence strategies and give recommendations for best observing strategies for time-lag estimation using photoRM, as well as to perform further tests on larger samples of observed light curves.


\section*{Acknowledgments}

This project is graciously supported by the 2021 LSST Corporation Enabling Science Call for Proposals (Grant Award \# 2021-11), and funding provided by Faculty of Mathematics University of Belgrade (the contract 451-03-9/2021-14/200104) and Astronomical Observatory (the contract 451-03-68/2020-14/200002), through the grants by the Ministry of Education, Science, and Technological Development of the Republic of Serbia. D.I. acknowledges the support of the Alexander von Humboldt Foundation.











\nocite{*}
\bibliography{Wiley-ASNA}%



\end{document}